# Matchmaking Semantic Based for Information System Interoperability

I Wayan Simri Wicaksana[1*]

[1] Gunadarma University, Jalan Margonda Raya 100, Depok, 16424, Indonesia

Unlike the traditional model of information pull, matchmaking is base on a cooperative partnership between information providers and consumers, assisted by an intelligent facilitator (the matchmaker). Refer to some experiments, the matchmaking to be most useful in two different ways: locating information sources or services that appear dynamically and notification of information changes. Effective information and services sharing in distributed such as P2P based environments raises many challenges, including discovery and localization of resources, exchange over heterogeneous sources, and query processing. One traditional approach for dealing with some of the above challenges is to create unified integrated schemas or services to combine the heterogeneous sources. This approach does not scale well when applied in dynamic distributed environments and has many drawbacks related to the large numbers of sources. The main issues in matchmaking are how to represent advertising and request, and how to calculate possibility matching between advertising and request. The advertising and request can represent data or services by using many model of representation. In this paper, we address an approach of matchmaking by considering semantic agreement between sources.

## 1. Introduction

Recently, research on information systems has increasingly focused on how to effectively manage and share data and services in heterogeneous distributed environments. Various sources can be accessed on-line in the web, including web pages, semi-structured documents (XML, RDF, etc.) and spatially referenced

Unlike the traditional model of information pull, matchmaking is base on a cooperative partnership between information providers and consumers, assisted by an intelligent facilitator (the matchmaker) [1]. Refer to his experiment, he said matchmaking to be most useful in two different ways: locating information sources or services that appear dynamically and notification of information changes.

. One traditional approach for dealing with some of the above challenges is to create unified integrated schemas or services to combine the heterogeneous sources. This approach does not scale well when applied in dynamic distributed environments and has many drawbacks related to the large numbers of sources. An alternative solution increasingly used in server oriented distributed environments is the semantic web, web services and ontology. One of component the approach is matchmaking

The main issues in matchmaking are how to represent advertising and request, and how to calculate possibility matching between advertising and request. The advertising and request can represent data or services by using many model of representation.

Earliest matchmakers based on KQML [2]. Similar approaches were deployed in SIMS and InfoSleuth. The matching process is carried through five progressive stages, going from classical Information Retrieval (IR) analysis of text to semantic match via Θ-subsumption. No ranking is presented but for what is called relaxed match, which basically refers again to a IR free text similarity measure.

A combination of semantic web or web services and Peer-to-Peer (P2P) technologies have potential to be an effective means for solving integration problems (e.g. data consistency, discovery, validation, etc.) due to their distributed nature and interoperability features. Semantic web and web services need to be able to communicate with each other which can be made possible by the use of P2P technology because of their: distributed nature, scalability, flexibility, manageability, deal with data or protocol and machine heterogeneity.

Sapkota [3] implemented P2P technology for web service discovery and matchmaking by employing web service modelling ontology (WSMO) as the underlying framework for describing both requests and service. He adapted hybrid P2P network is similar to the super peer based model. Peers are described semantically and the super peer is a relative concept. Clustering is implemented based on similarity of services. each cluster consists of at least one super peer and it is maintained and managed by one of the super peer. The matchmaker of Sapkota is centralized at super peer, it is will overloaded and single failure problem of super peer. Our approach will distribute to peer as provider to execute matchmaking, and result directly send to peer as request. In this paper, we are interested to look at implementation of matchmaking to create half and full agreement among the parties. Result of agreement can be used for peer-to-peer (P2P) fro discovery and query process

Our focus in this paper is on P2P by using semantic agreement. We propose an approach based on a P2P for data and services interoperability of information sources that aims to combine the advantages of semantic agreement and peer-to-peer systems. It is based on a hybrid P2P architecture consisting of two types of peers. The main task of the super peer is to register active peers and the meta data used to describe their contexts. Peers can be data request or provider. Main our contribution is how to create and implement peer semantic agreement for discovery process. Process calculation of semantic similarity based on current available approaches.

The paper is organized as follows. Section 2 presents the peer agreement based semantic approach. Discussion will be addressed at section 3. And finally section 4 concludes the paper

## 2. Peer Agreement Approach

Publish, Request and Bind process as figure 1, it has minimal requirement for the functionalities of information or service interoperability are:

- Publish: providers can publish their description of the features of the data or services which are providing. In our approach the publishing will introduce with preprocessing which called agreement (half agreement). The agreement (half agreement) is result of matchmaking/mapping phase 1 between export schema of a peer to common ontology of a super peer. Developing of agreement based on current approach (label matching and external structure comparison.
- Request: a peer send a request to find or locate sources for his query to search a relevant advertisement among the currently available ones. Firstly the peer search through meta data at super peer for candidates sources. From the first step, list of candidate will be created. The peer can broadcast his half agreement to candidates peers and calculate similarity/matchmaking phase 2 between half agreement of peer as request and peers as provider. Result of calculation can be exact agree, similar and non similar. The sources peer (interest parties) can be decided after this ste
- Bind: interest parties can create mapping composition based on their half agreement. The interest parties will create a mapping composition which can be used to exchange data or services.

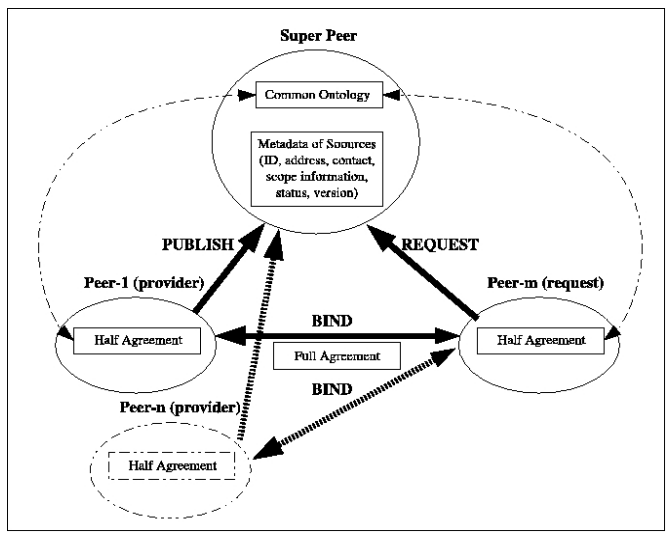

Figure 1. Publish, Request and Bind in P2P

The calculation of semantic similarity between concepts is the first step for creating agreements between the ontology and the contents of the other peers. Each concept can be represented as a hierarchy of terminological labels which contains some structural or semantic information.

The first matching is the label associated with the concepts to determine concepts that are related. A label has a part value of semantic which presented at taxonomy model such as WordNet. The result of this matching process can be fine tuned by considering both the internal structure and the external structure of the concept represented by a label. The internal structure of a concept consists of its directly linked attributes while the external structure takes into account the position of a concept in a hierarchy.

## 3. Discussion

We tested the approach proposed here using the three domain from different papers. There are Bishr [4] for transportation domain, Cruz [5] for publication domain and Rahm [6] for business domain.

Calculation of Label Matching used WordNet v2.1, WordNet::QueryData v1.40 and WordNet::Similarity v1.03. Calculation of External structure used set of superclass, we can not used leave method because the schemas are too simple. During the experiment we adjust value of some threshold value to get the optimal result.

We evaluate the performance of our approach based on three measures: precision, recall and F-measure. Given (1) the number of match M determined by a related paper, (2) the number of correct mapping R compare between (1) to calculation based on our approach, and (3) the number of incorrect mapping C selected by our precess. We compute the recall ration as $R=C/N$, the precision ration as $P=R/(R+C)$, and the F-measure as $F=2/(1/R +1/P)$ . We report all these values as percentages.

The result of performance of experiment can be seen figure 2 with threshold value of label matching is 0.9, threshold value of external value is 0.49, and threshold value of confidence value is 0.75. The result has not yet used user feedback.

External structure can enhance the result, for example from Rahm's schema value of *BillTo*:Name is different with *ShipTo*:Name. If the procedure just utilize label matching, the result is same, but by adding external structure the value of confidence can differ around 15%.

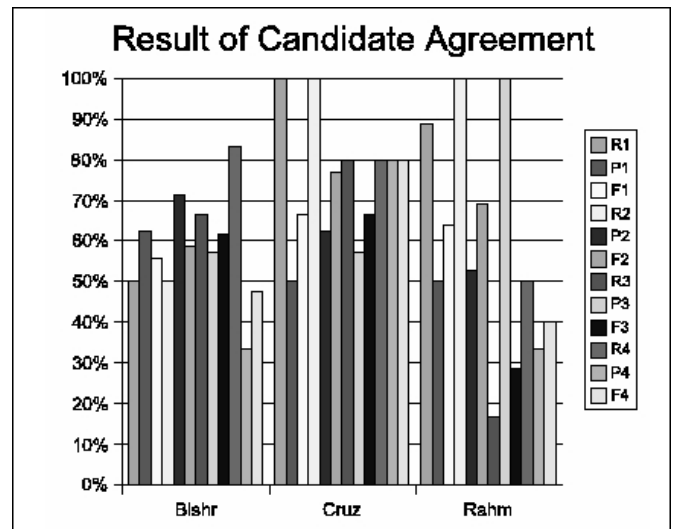

Figure 2. Result of Experiment

Unsatisfied of result is depend on 'completeness' of linguistic (WordNet) taxonomy. Previously, we have evaluated two previous version of WordNet, the result is the current version is better then previous versions. The low value of performance for Bishr at CO-S1 because there is no word of interstate and pedestrian at WordNet. The very low of performance for Rahm at CO-S2 because the depth of S2 is 'flat', so enhancement by using external structure can not be implemented.

From above result, our approach can be implemented to create half agreement which can utilized for publish, request, full agreement, and query. The result will depend on taxonomy/ontology for label matching and common ontology. The result can be enhanced by bring the power of OWL through reasoning process.

## 4. Conclusion

XMLS, RDFS and OWL and other ontology developments offer facility to enrich semantic description at P2P environment. We proposed a semantic agreement approach based on concept similarity values that take into account the place of a concept in a hierarchy and its structure consisting of directly linked properties and concepts; We have described general processing steps based on the proposed approach. Result of agreement (set of agreement unit) can be utilized for matchmaking in discovery resources, mapping of concept and handling of query.

We will focus on finalizing the architecture and prototype system to enhance negotiations between provider peer and a peer to improve the matchmaking, query response and reduce performance load of the peers.